\documentclass[a4paper]{article}
\usepackage{graphicx}
\usepackage{amsmath}

\begin{document}

\begin{center}
{\bf \Large
How political parties adjust to fixed voter opinions 
}\\[5mm]

{\large
Krzysztof Ku{\l}akowski
}\\[3mm]

{\em
Faculty of Physics and Applied Computer Science,
AGH University of Science and Technology,
al. Mickiewicza 30, PL-30059 Krak\'ow, Poland
}

\bigskip
{\tt  kulakowski@novell.ftj.agh.edu.pl}

\bigskip
\today
\end{center}

\begin{abstract}
We propose a new version of the spatial model of voting. Platforms of five
parties are evolving in a two-dimensional landscape of political issues so
as to get maximal numbers of voters. For a Gaussian landscape the evolution 
leads to a spatially symmetric state, where the platform centers form a pentagon 
around the Gaussian peak. For a bimodal landscape the platforms located at 
different peaks get different numbers of voters.
\end{abstract}



\section{Introduction}

Dynamics of public opinion is a central subject in political sciences. As votes can be described 
with numbers, research in this field 
belongs at least partially to the behavioral tradition in America or to the sociophysics 
in Europe. Indeed, opinion dynamics attracts 
attention of several authors in the Old Continent \cite{sznajd,krheg,deff,sta,sta2}. In these 
approaches, the process considered is that voters, convinced by other voters, change their 
opinions about parties. However, there has been also an opposite point of view, established in 
literature for 50 years \cite{downs}. According 
to this point, parties adopt political platforms in order to to get maximal number of voters. Although 
it is clear that in reality both processes occur,
it seems advisable to investigate the latter separately, as if opinions of voters remain 
constant in time. Such a search is our purpose here. 

Actually, there is at least one argument that the variations of political platforms to meet the 
voter's preferences are quicker than the changes
of voter's preferences. This argument is as follows: a standard voter is not economically 
motivated to optimize his performance. It is clear that
one vote cannot change a political landscape. Then our political preferences are based rather
 on an identification with a given politician
than on acceptance of his program. Programs are long, complicated and devious, whereas people 
can be qualified as fine or not in seconds
\cite{buss}. As a result, many vote for candidates who look good on TV. On the contrary, politicians 
are strongly motivated to fight for voters. 
There, the difference between success and failure is equivalent to the difference between being 
Prime Minister and being unemployed.  
In this aspect, politicians can then be expected to be much smarter, better informed and quicker 
than voters. If this is so, the variations of 
platforms can be described with an assumption that the preferences of voters are constant in time. 
This is a sort of adiabatic approximation. In fact, both processes occur: the platforms move and 
the voters change their opinion, but in many cases the characteristic time of the latter is longer.
Once the platforms are established, most of voters have no choice but to vote for a platform 
prepared for them: workers for the left, bussiness for the right,
intellectuals for professors, young for greens etc. The coupling between well-defined platforms 
and clusterized groups of voters has an additional feedback formed by media: every reader finds a 
newspaper where things are presented according to his own opinion. To be precise, a platform should 
be understood as a set of issues which can serve as criteria for voters. A choice of these issues 
does depend on tradition and history. However, final output of a candidate appears to depend also on 
his/her charisma, age, sex, height and health which we are well-trained by the Darwinian evolution to evaluate 
\cite{buss}. 

In this perspective, the game between politicians and voters is no more equivalent to a time 
evolution of the statistical distribution of opinions
on static issues. It is close rather to a deterministic search for herds of voters, unable 
to change their opinions. These herds form a 
political landscape in a multidimensional space of issues \cite{riker}. The picture is known as 
spatial voting model. The deterministic character
of the time evolution can however be relaxed by the incomplete knowledge on the public opinion. 
Indeed, much money is paid by governments
to recognize the voters' response for issues which could or could not be a basis of a winning 
platform \cite{cbos}. Here we adopt the approach
of Kollman, Miller and Page \cite{mil1,mil2}, who simulated the evolution of platforms in a 
given landscape. In these works, the incomplete knowledge on the 
landscape was reflected by the time evolution rules, determined by the landscape only at the 
actual position of the platform. Here we use the same locality principle. On the other hand, 
we feel to continue the sociophysical tradition, asking for the probability distribution of 
votes \cite{sznajd,sta2}. The aim of this paper is to investigate
this distribution in a given landscape. Namely, we ask if there is any connection between the 
distribution of votes and the shape of the landscape.
  
In Section II the model is explained. The results are described in Section III and discussed in 
Section IV. Final conclusions close the text.

\section{The model}

In computer simulations, the incompleteness of knowledge was reflected \cite{mil1,mil2} by using 
one of three approaches: random search, 
local hill-climbing and genetic algorithm. As the results of these approaches are qualitatively 
the same, we feel free to use one of them, 
namely the hill-climbing algorithm. The model space of issues is limited here to two dimensions, 
$x$ and $y$. The criterion of selection of issues is that they should have a 
discriminative power. For example, slavery does not fulfil 
this criterion. On the contrary, this discriminative power cannot be too large; a woman who 
wants her husband to be Prime Minister cannot gather a party around this postulate. Still,  
a rich spectrum of possible between these two extremes.

Having the axes, one should be able to construct the landscape. Here again we encounter another 
eternal problem in social sciences: the scale. 
As it is known from the utility theory, scales do depend on the respondent \cite{stra}, what 
makes the construction subjective. Various solutions
of the problem can be found in \cite{oxford}. Here we intend to postulate that a landscape of an 
unbiased issue should be close to a Gaussian function,
just by nature of statistics. By unbiased we mean that {\it i)} no abrupt changes of the opinion 
happened recently, {\it ii)} people are not personally
engaged into an issue. If they are engaged, bimodal distributions are likely to appear \cite{kk}. 

Initial positions of the platforms are selected randomly, with uniform 
distribution. The number of platforms is arbitrary, but this 
choice is supported by some 
common sense. Parties which get small amounts of votes do not enter into parliaments in many 
countries. Moreover, their results in our simulation
would be probably distorted by statistical errors of the order of their yields. The time 
evolution is governed by the principle of hill-climbing: if a party can 
get more votes by a shift of the position of its platform, the shift is done. We note that 
this algorithm was checked in \cite{mil1} to produce similar results as
the random-search algorithm and the genetic algorithm. The length of steps in the space of 
issues is arbitrary, but small with respect to a characteristic 
length of the landscape variation. Our algorithm is equivalent to a 
differential equation

\begin{equation}
\frac{\delta \mathbf{y}_i}{\delta t}=\nabla _y w_i(\mathbf{y}) 
\end{equation}
where $\mathbf{w}=(w_1,..,w_5)$ is the number of votes gained by $i-th$ party at position 
$\mathbf{y}_i$ in two-dimensional space of issues. The number 
of votes of $i-th$ party is calculated from its position in the space of issues, 

\begin{equation}
w_i=\int d\mathbf{x} \rho(\mathbf{x})g(\mathbf{y_i}-\mathbf{x})[1-\frac{1}{N}\sum_j g(\mathbf{y_j}-\mathbf{x})]
\end{equation}
where the function $g(\mathbf{x})$ describes the profile of votes as dependent on the position 
of the platform. Here it is selected to be also Gaussian, with 
the width $\sigma $ set as $2^{-1/2}$. The second term under the integral 
describes the interaction between the parties, which is
repulsive; it can be more beneficial for a party to explore voters in an area where other parties 
are not active, even if the number of voters is somewhat smaller there.
From the point of view of a physicist, the defined system is analogous to five interacting 
overdamped particles, looking for local equilibria in a potential minimum.
The potential is the landscape with inverted sign. The resulting set of equations of motion 
for $i=1,...,5$ is

\begin{equation}
\frac{d\mathbf{y}_i}{dt}=-\frac{2\mathbf{y}_i}{1+2\sigma^2}I(i)+
\sum_j\frac{2\mathbf{y}_i+4\sigma^2(\mathbf{y}_i-\mathbf{y}_j)}{N(1+4\sigma^2)}J(i,j)
\end{equation}
where $I(i),J(i,j)$ are scalar quantities

\begin{equation}
I(i)=\frac{1}{\pi(1+2\sigma^2)}\exp\Big(-\frac{\mathbf{y}_i^2}{1+2\sigma^2}\Big)
\end{equation}
and

\begin{equation}
J(i,j)=\frac{1}{\pi(1+4\sigma^2)}\exp\Big(-\frac{\mathbf{y}_i^2+\mathbf{y}_j^2+2\sigma^2(\mathbf{y}_i-\mathbf{y}_j)^2}{1+4\sigma^2}\Big)
\end{equation}
and $N=5$ is the number of parties. The term with $J(i,j)$ describes 
the repulsion between parties $i$ and $j$. We keep $J(i,i)=0$. Equation (3) is solved numerically.
In general, Eq. 1 reduces to a differential equation, provided that the product of functions
under the integral (2) can be approximated by a linear combination of products of Gaussian functions and polynomial 
functions.

\section{Results and discussion}

First we consider the landscape which is a single Gaussian function with its maximum at the 
coordination centre.
We performed the calculations for several sets of initial positions of the platforms in the 
two-dimensional
space of issues. For large values of $\sigma$, the emerging result is always the same: 
the centers of the platforms
tend to equidistant positions on a circle, formed around the peak of the Gaussian peak of the 
landscape. Example of the trajectories is shown in Fig. 1. Even if the initial position
of one party is on the top of the peak, i.e. in the center of coordinates, this party gets 
down the peak and finally is placed on the circle. In stable equilibrium, the yields
$w_i$ of all the parties become equal. This kind of 
symmetry should appear for any number of parties; we checked that it is true for $N=2$.

In principle, it could be expected that one party,
initially closest to the centre, will be able to fix its platform there before the other 
parties. On the contrary to this expectations,  a central platform placed at the peak
gets down and moves to a position equivalent to those of other platforms. The 
memory of the initial state is lost except the angular coordinates of the parties. In Fig.2
we show final positions of the platforms, reached from several random initial positions.

If the width of the landscape peak $\sigma$ is small enough, it becomes worthwhile for the
parties to occupy the top of the peak even if shared with platforms of other parties. We can apply
the stability analysis to investigate the stability of the situation when all 
platforms are situated at the peak top. For two parties, the result is analytical: for 
$\sigma>((1+2^{1/2})/2)^{1/2}\approx 1.1$, the point $\mathbf{y}_i=\mathbf{0}$ is not stable.
This means, that the coexistence at the top is not fruitful.
For five parties, the critical value of $\sigma$ is about $1.18$. However, even fairly below
this value the time evolution of the platforms is very slow near the top, and the above 
stability is hard to be evaluated from the numerical solution. 

%% ----------------------------------------------------------------------------
\begin{figure}
\begin{center}
\includegraphics[angle=-90,width=.8\textwidth]{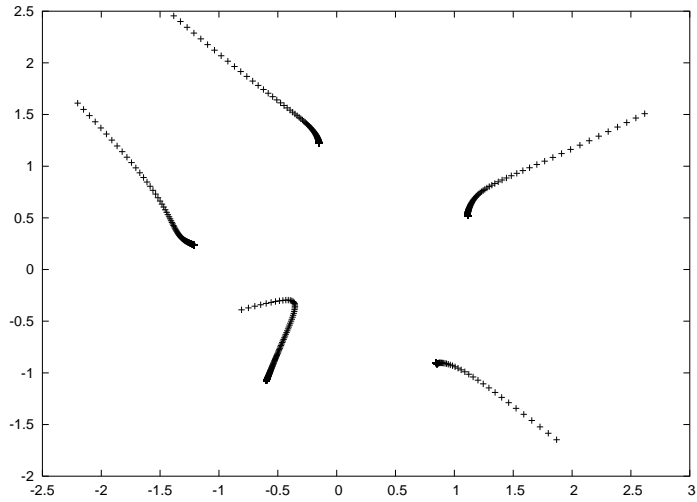}
\caption{Trajectories of platforms of five parties, starting from random initial positions.
The density of points increases with time, because the velocity of platforms decreases. 
This reveals the direction of the trajectories,
which is generally to the center. However, one of them (in the center of the lower part
of the figure) changes the direction, repulsed by the others.}
\label{fig1}
\end{center}
\end{figure}
%% ----------------------------------------------------------------------------

%% ----------------------------------------------------------------------------
\begin{figure}
\begin{center}
\includegraphics[angle=-90,width=.8\textwidth]{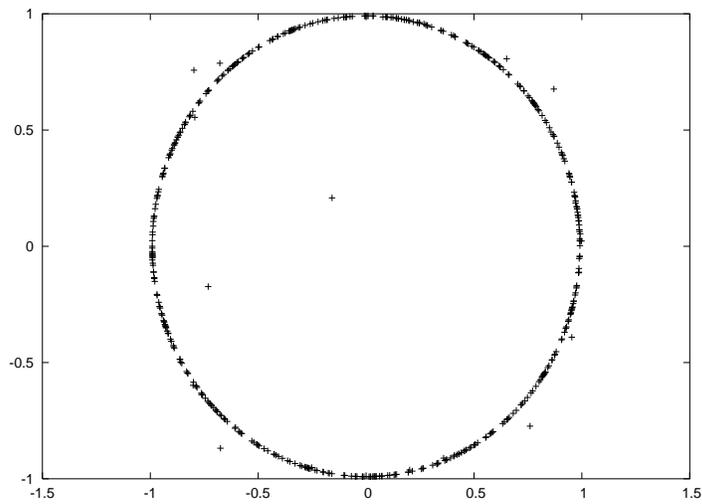}
\caption{Positions of five platforms after some time, averaged over 100 random initial positions.}
\label{fig2}
\end{center}
\end{figure}
%% ----------------------------------------------------------------------------

%% ----------------------------------------------------------------------------
\begin{figure}
\begin{center}
a)\includegraphics[angle=-90,width=.5\textwidth]{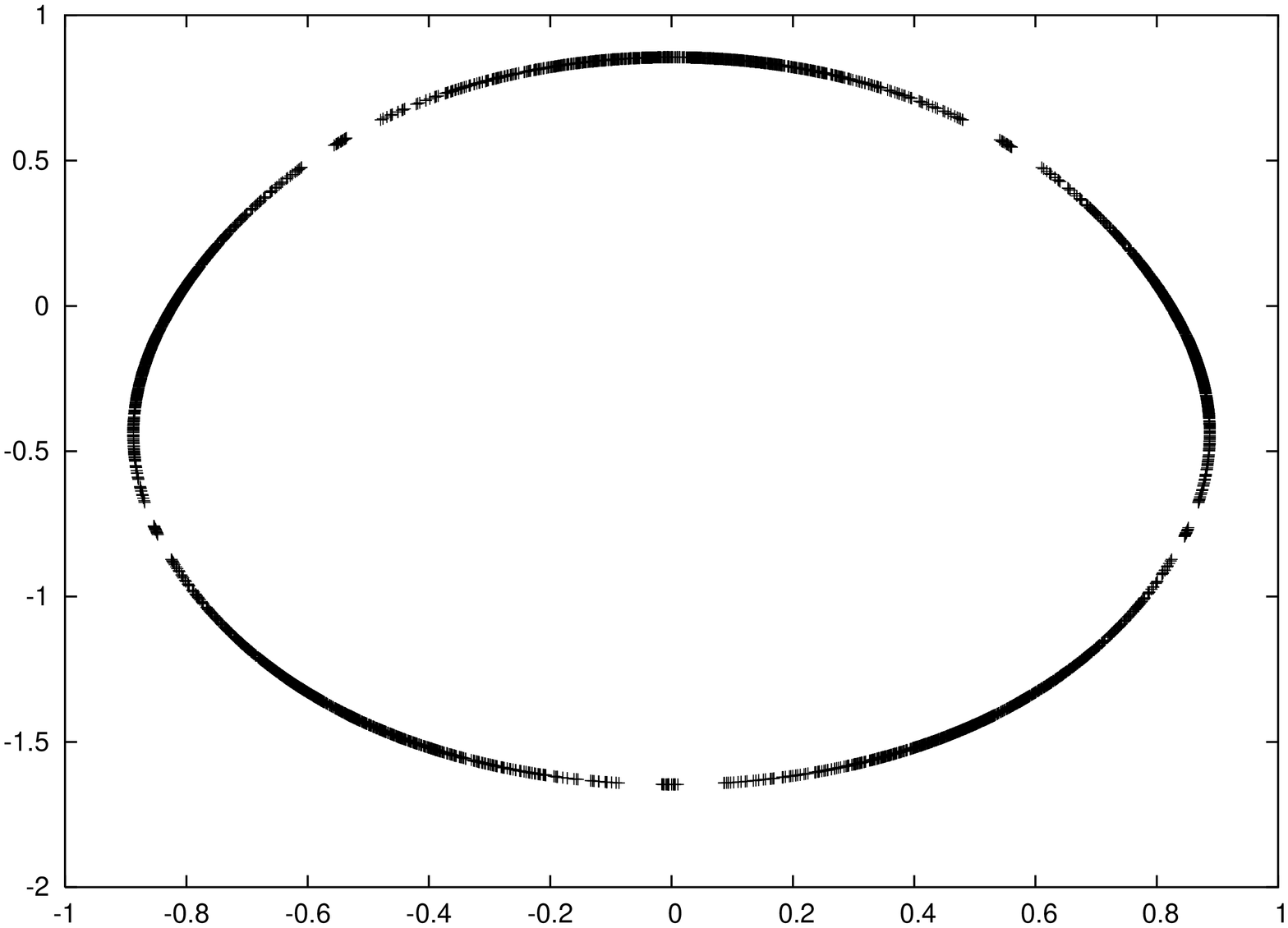}
b)\includegraphics[angle=-90,width=.5\textwidth]{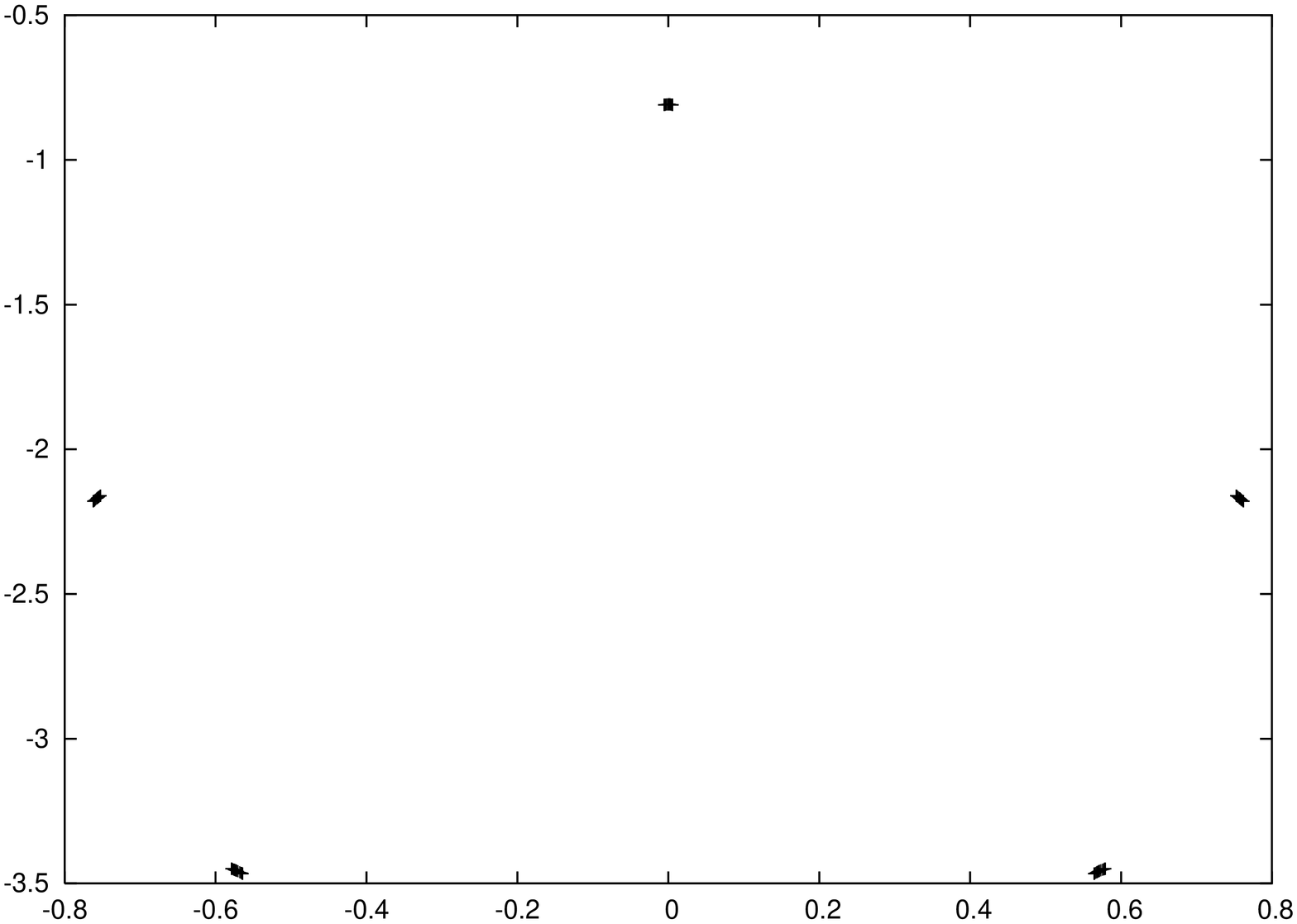}
c)\includegraphics[angle=-90,width=.5\textwidth]{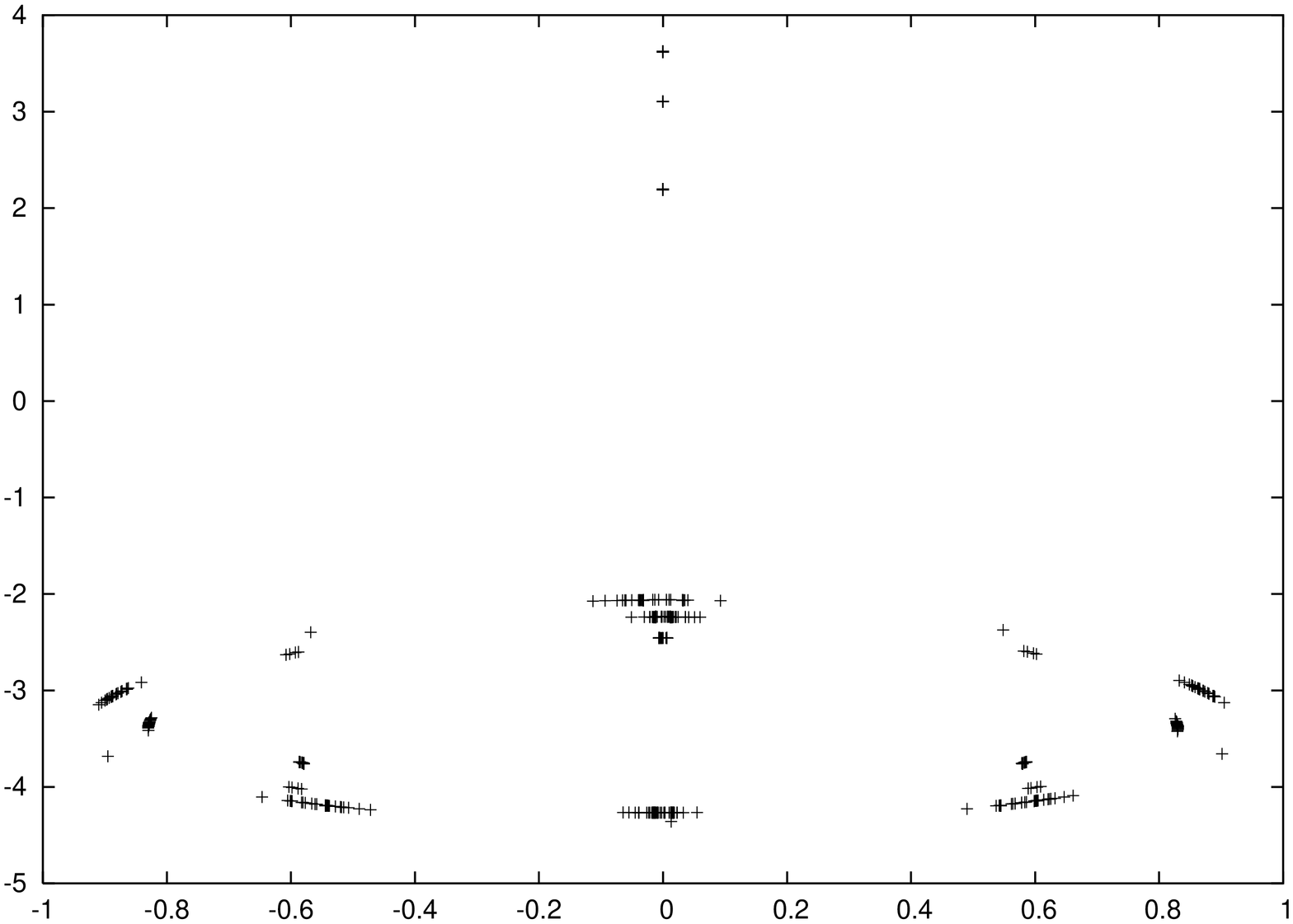}
d)\includegraphics[angle=-90,width=.5\textwidth]{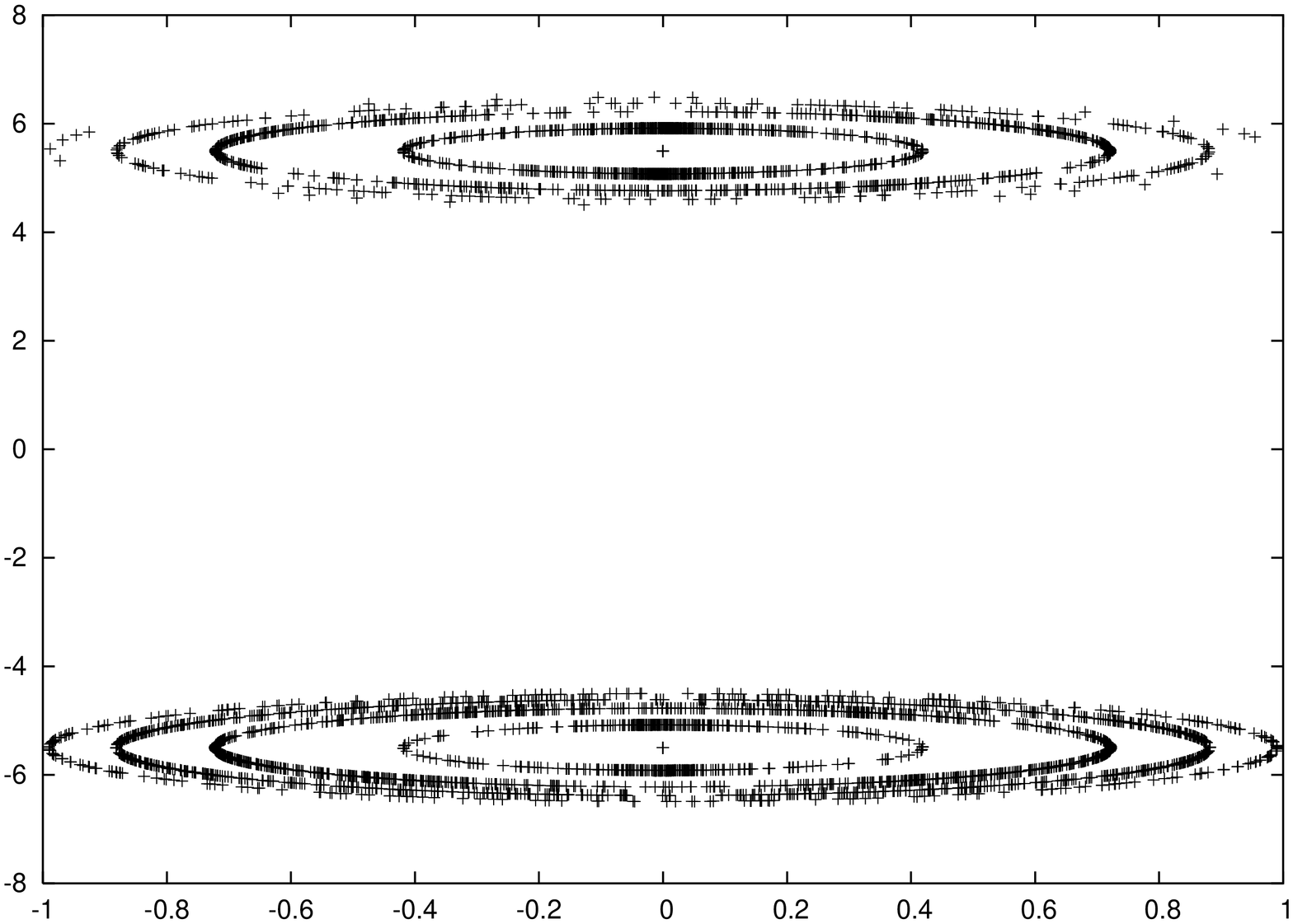}
\caption{Positions of five platforms after some time, averaged over 1000 random initial positions
for a) $c=1.5$, b) $c=3.0$, c) $c=3.5$, d) $c=5.5$.}
\label{fig3}
\end{center}
\end{figure}
%% ---------------------------------------------------------------------------

It is clear that the obtained circular symmetry must vanish if the landscape is not symmetric.
As it was recognized in \cite{mil2}, the ability of a platform to get an optimal position
decreases with the landscape ruggedness. In particular, for a bimodal landscape it is obvious 
that the hill-climbing algorithm traps some platforms at a peak which is maybe more 
occupied and therefore less favorable. 

Now we consider a landscape formed from two Gaussian peaks, placed in equal distances from
the coordination centre. Their coordinates are $(0,c)$ and $(0,-c)$, and the latter is $3/2$ 
times higher. As chosen previously, $\sigma =2.5$ for both peaks. In Fig.3 we show how the 
final positions of five platform centers are distributed for various values of $c$. As we see, 
for short distance $2c$ between the peaks all parties are close to the higher peak. In this case,
their yields $w_i$ are equal. As $c$ increases, we observe some kind of phase locking;
a platform is located between the peaks and the repulsion between platforms leads them to fixed
positions on some curve around the peak where $w_i=const$. Finally, when $c$ is large enough,
the peaks can be considered as independent. In this case the peak selected by a platform does 
depend on the initial position of all the platforms. In principle, all partitions are possible,
i.e. $(0-5)$, $(1-4)$, $(2-3)$, $(3-2)$, $(4-1)$ and $(5-0)$. The weights of these partitions
depend on the shape of the landscape, but it may be approximated by the binomial 
distribution. The numbers of voters $w_i$ do depend on the partition. It is best for a party to
be only one occupying the maximum, even if this is the lower one.

\section{Discussion}

Some conclusions drawn from our results are at least not contradictive
with a common experience.  First, equilibrium positions of the platforms
are to be in maximal possible distance. This makes an accordance between parties generically
difficult, even if they are close to each other in their programs. Second, the strongest hostility can be expected 
between parties with neighboring platforms, because they fight for voters. Third, isolated maxima
of the electoral landscape are expected to be willingly occupied even if these maxima
are small. This is our contribution to an interpretation of extremist parties, which appear to be 
good ecological niches for some politicians. As a rule, the bosses of these parties are 
authoritary, as they are not forced to make compromises with neighbors - they have none.
Fourth, the emerging picture 
is a convenient basis to investigate the response of parties for evolution of the electoral
landscape. In particular, suppose that we take into account an increasing disappointment 
of voters with a ruling party. Their program is not executed or not fully executed, affairs
disgrace their government and initial hope that their electoral victory will push the country
into a prosperous future has no more support. To introduce these known facts to the spatial 
model, it is enough to reduce gradually the function $g$ of the ruling party, until its supremacy 
is lost. The effect is known as political pendulum. After several cycles, the process leads
to a fragmentation of political scene, until new issues appear. 

To conclude, in the spatial model of voting the positions
of the political platforms is a part of the game, and it has not much to do with historical
tradition of the parties. We note that this result cannot be obtained in a one-dimensional model,
where the repulsion between parties prevents them to profit the same groups of voters. In a sense,
our model could be applied to a problem of division of territories with herds of cattle 
between shepherds, or groups of buyers between companies. In all these problems, repulsion
between shepherds (or companies or platforms) is a natural consequence of deficience of resources
in areas where two owners can met. Our results indicate that the positions of platforms
display a kind of a collective optimization, where a supremacy of one party is unstable. Politically,
the emerging system can be compared to an oligarchy, where the influence of each local ruler
is taken into account by its neighbors.

We should note that there are also other politicians and parties who - for various reasons -
do not try to get more votes. Obviously, their performance cannot be captured with the above 
description. However, it is only rarely that we can see them as winners of an election. 
\bigskip

{\bf Acknowledgements.} The author is grateful to Dietrich Stauffer for a stimulating discussion,
and to the Organizers of the 8-th Granada Seminar for their kind hospitality.


\begin{thebibliography}{99}
\bibitem{cbos} Reports of opinions of the author's compatriots on the Polish Government are published each month by the Public Opinion Research Center (CBOS).
\end{thebibliography}
\end{document}